\documentclass[aps,prx,twocolumn,floatfix,longbibliography,nofootinbib,superscriptaddress]{revtex4-1}
\usepackage[utf8]{inputenc}
\usepackage{natbib}
\usepackage{graphicx}
\usepackage{xcolor}
\usepackage[tbtags]{amsmath}
\usepackage[colorlinks, linkcolor=blue]{hyperref}
\hypersetup{colorlinks,allcolors=black}
\usepackage{amssymb}
\usepackage{amsmath}
\usepackage{gensymb}
\usepackage{float}
\usepackage{amsmath}
\usepackage{tabularx,graphicx}
\usepackage{epstopdf}
\usepackage{latexsym}
\usepackage{color, colortbl}
\usepackage{psfrag}
\usepackage{bbm}
\usepackage{bm}
\usepackage{titlesec}
\usepackage{dsfont}
\usepackage{feynmp}
\usepackage{slashed}
\usepackage{multirow}
\usepackage[normalem]{ulem}
\renewcommand{\vec}[1]{\boldsymbol{#1}}

\newcommand{\bcen}{\begin{center}}
\newcommand{\ecen}{\end{center}}
\newcommand{\btab}{\begin{tabular}}
\newcommand{\etab}{\end{tabular}}
\newcommand{\bdes}{\begin{description}}
\newcommand{\edes}{\end{description}}

\newcommand{\beq}{\begin{equation}}
\newcommand{\eeq}{\end{equation}}
\newcommand{\bea}{\begin{eqnarray}}
\newcommand{\eea}{\end{eqnarray}}

\newcommand{\bary}{\begin{array}}
\newcommand{\eary}{\end{array}}
\newcommand{\benum}{\begin{enumerate}}
\newcommand{\eenum}{\end{enumerate}}
\newcommand{\bitem}{\begin{itemize}}
\newcommand{\eitem}{\end{itemize}}

\renewcommand{\vec}[1]{\boldsymbol{#1}}
\newcommand{\br} { \boldsymbol{r}}

\newcommand{\B} {{\mathcal{B}}}

\newcommand{\Fig}[1]{Fig.~\ref{#1}}

\makeatletter

\newcommand{\Rmnum}[1]{\expandafter\@slowromancap\romannumeral #1@}
\makeatother

\newcommand{\tn}{\textnormal}
\newcommand{\nn}{\nonumber}

\begin{document}

\title{Fractionalization and topology in amorphous electronic solids}
\author{Sunghoon Kim}
\affiliation{Department of Physics, Cornell University, Ithaca NY 14853, USA}
\author{Adhip Agarwala}
\affiliation{International Center for Theoretical Sciences, Bangalore 560089, India}
\affiliation{Max-Planck Institute for the Physics of  Complex Systems, Nöthnitzer straße 38, Dresden 01187, Germany}
\affiliation{Department of Physics, Indian Institute of Technology, Kanpur 208016, India}

\author{Debanjan Chowdhury}
\affiliation{Department of Physics, Cornell University, Ithaca NY 14853, USA}

\date{\today}

\begin{abstract}
Band-topology is traditionally analyzed in terms of gauge-invariant observables associated with crystalline Bloch wavefunctions. Recent work has demonstrated that many of the free fermion topological characteristics survive even in an amorphous setting. In this work, we extend these studies to incorporate the effect of strong repulsive interactions on the fate of topology and other correlation induced phenomena. Using a parton-based mean-field approach, we obtain the interacting phase diagram for an electronic two-orbital model with tunable topology in a two dimensional amorphous network. In addition to the (non-)topological phases that are adiabatically connected to the free fermion limit, we find a number of strongly interacting amorphous analogs of crystalline Mott insulating phases with non-trivial chiral neutral edge modes, and a fractionalized Anderson insulating phase. The amorphous networks thus provide a new playground for studying a plethora of exotic states of matter, and their glassy dynamics, due to the combined effects of non-trivial topology, disorder, and strong interactions.

\end{abstract}

\maketitle

Topological band theory has made a profound impact on our fundamental understanding, and in the experimental search, of weakly correlated phases of crystalline electronic solids \cite{Ludwig_PS_2015,Chiu_RMP_2016,Hasan_RMP_2010, Qi_RMP_2011}. A useful starting point in this endeavor is to consider crystals with perfect translational symmetry, and to include perturbative corrections due to the effects of weak disorder \cite{Sheng_PRL_2006, Xu_PRB_2006,Wu_PRL_2006,Onoda_PRL_2007, Liu_PRL_2009, Medhi_PRB_2012, Fulga_PRB_2012}. However, translational symmetry is not a prerequisite for realizing non-interacting topological phases, even in the absence of a well-defined momentum, an associated electronic ``bandstructure" and the very notion of a Brillouin zone \cite{Agarwala_PRL_2017, Mitchell_NP_2018, poyhonen2018amorphous, Chern_EPL_2019, costa2019toward, Yang_PRL_2019, spring2021amorphous,Agarwala_PRR_2020}. Interestingly, experiments on a candidate topological material \cite{2019arXiv191013412C} and a photonic system \cite{Zhou2020photonic} provide promising evidence for the existence of topological phases in amorphous settings. On the theoretical front, the study of realistic models and better computational techniques continue to advance our understanding of such amorphous topological phases of free fermions \cite{marsal2020topological,Ivaki_PRR_2020,Focassio_PRB_2021}.

\begin{figure}
	\centering
	\includegraphics[width=0.95\linewidth]{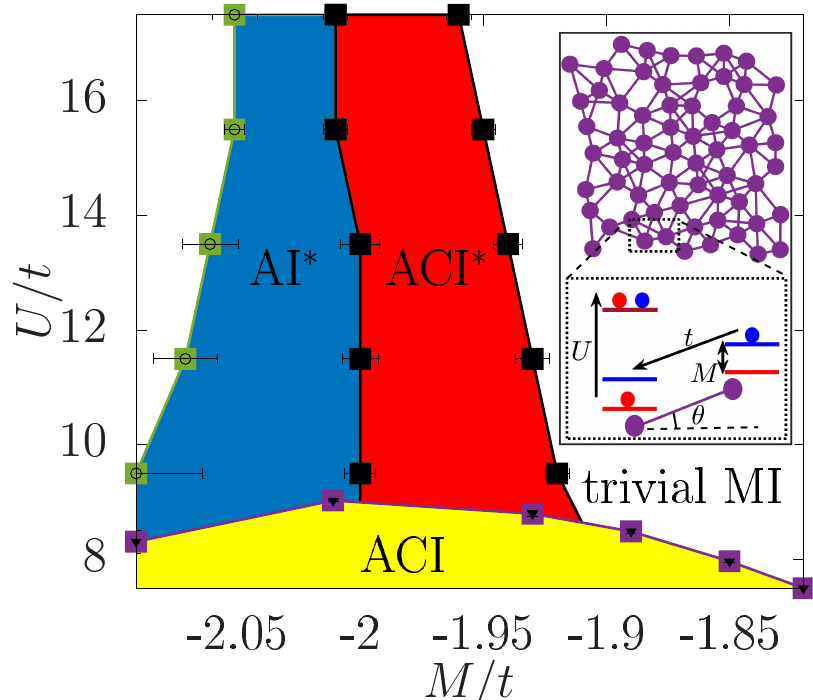}
	\caption{Theoretical phase diagram at half-filling as a function of $M/t$ and $U/t$ for system of size $16\times 16$. The four distinct phases include a trivial electronic Mott insulator (MI), an amorphous Chern Insulator (ACI), a fractionalized Anderson insulator (AI$^*$), and a fractionalized amorphous Chern insulator (ACI$^*$), respectively. The inset denotes a typical amorphous network and the energy scales associated with hopping ($t$), orbital splitting ($M$) and on-site repulsion ($U$).}
\label{fig:phasediagram}
\end{figure}

In this letter, we address a question of fundamental interest that has not received as much attention, except when in a  crystalline setting \cite{Maciejko_PRL_2010,Rachel_PRB_2010, pesin2010mott, He_PRB_2011, Swingle_PRB_2011, Maciejko_NatPhys_2015, rachel2018interacting, Hickey_PRB_2015, Roy_PRB_2016}, namely on the effects of strong local interactions on the fate of the amorphous topological phases. The combined effects of strong interactions, topology and disorder (manifest in the amorphous structure of the underlying network) can lead to a plethora of new phases. To study the interacting problem, we employ and adapt a parton mean-field theory \cite{Florens_PRB_2004,Zhao_PRB_2007, Potter_PRL_2012} in this new amorphous setting and obtain the theoretical phase-diagram for an interacting two-dimensional electronic model as a function of the strength of interactions and tunable topology. At fixed electron-filling and as a function of increasing interaction strength, we find evidence of transitions into phases with deconfined fractionalized excitations, that can either support neutral edge modes (as in an amorphous Chern insulator), or display disorder-induced localization (as in an Anderson insulator) accompanied by a local inhomogeneous distribution of topological ``puddles". We use a number of quantitative diagnostics for disentangling the effects of disorder and topology in order to sharpen the properties of the resulting phases and their regions of stability with increasing system sizes. While our study of a simplified theoretical model is not directly related to a realistic model of a quantum material, recent experimental work in twisted bilayer graphene \cite{matbg} suggests that the notion of a ``local" Chern number and a ``mosaic" of distinct topological patches can emerge in the presence of long-wavelength inhomogeneities in a strongly correlated topological system. Our theoretical study highlights the incredibly rich playground provided by seemingly ``disordered" systems as an exciting new platform for realizing topological and fractionalized phases of matter.

{\it Model.-} We define a two-dimensional (spinless) electronic model with local interactions, residing on an amorphous network with two orbitals (e.g. $s$ and $p$) per site; see Fig.~\ref{fig:phasediagram}-inset. The Hamiltonian is of the form,
\begin{subequations}
\bea
H &=& H_0 + H_{\tn{int}}, \label{eq:Ham}\\
H_0 &=& - \sum_{\substack{i\neq j\\ \alpha,\beta}} \bigg[t_{\alpha\beta}(\br_{ij}) c_{i,\alpha}^\dagger c^{\phantom\dagger}_{j,\beta} + \tn{h.c.}\bigg] + \sum_{i, \alpha, \beta} \epsilon_{\alpha \beta} c^\dagger_{i,\alpha} c^{\phantom\dagger}_{i,\beta},   \nn\\ \\
H_{\tn{int}} &=& \frac{U}{2} \sum_i\hat{N}_i(\hat{N}_i-1),
\eea
\end{subequations}
where $c_{i,\alpha}^\dagger (c^{\phantom\dagger}_{i,\alpha})$ is an electron creation (annihilation) operator with orbital index $\alpha$ at site $i$ and $t_{\alpha\beta}(\br_{ij})$ represents the tunneling matrix element between sites $i,j$ separated by distance $r=|\br_{ij}|$. The on-site single-particle and interaction energies are given by $\epsilon_{\alpha\beta}$ and $U$, respectively, where $\hat{N}_i = \sum_{\alpha} c^\dagger_{i,\alpha}c^{\phantom\dagger}_{i,\alpha}$. We assume that the tunneling matrix elements have a {\it statistical} rotational symmetry with an exponential fall-off with distance over a characteristic scale, $a_B$, and vanish beyond $r=R$, i.e. $t_{\alpha\beta}(\br) = t(r) T_{\alpha \beta} (\hat{\br})$ where $t(r) = C \Theta(R-r)\exp(-r/a_B)$ and  $C$ is a constant chosen such that $t(r=a_B)=1$. All other energy scales are scaled with respect to this unit scale. The orbital dependence of the single-particle energy and tunneling matrix element is given by,
\begin{subequations}
\bea
\epsilon_{\alpha \beta}&=& \begin{pmatrix}
2t+M  &    0  \\
0 & -(2t+M) 
\end{pmatrix},\\
T_{\alpha \beta}(\hat{\br}) &=&
\frac{1}{2}\begin{pmatrix}
-1  & -ie^{-i\theta} \\
-ie^{i\theta} &   1
\end{pmatrix},
\eea
\end{subequations}
where $\theta$ is the angle subtended between a reference axis and the vector $\hat{r}$ connecting sites $i$ and $j$ (see inset of Fig.~\ref{fig:phasediagram}), and $M$ modulates the on-site energy of the orbitals. The sites that make up the amorphous graph are constructed randomly, but with a minimal exclusion distance, $r_{\tn{min}}$. 

In the non-interacting limit ($U=0$), $H$ belongs to class A of the ten-fold classification with broken time-reversal symmetry, and where previous work has pointed out the existence of free fermion topological phases with robust edge states as a function of varying $M, a_B$  \cite{Agarwala_PRL_2017}. These phases exists over a large window of $a_B,r_{\tn{min}}$ \cite{SI}, and it is straightforward to generalize these results to other classes of the ten-fold way. In the remainder of this paper, we obtain the phase diagram of $H$ as a function of $U/t,M/t$, where the complex interplay of interactions, topology and the amorphous nature of the underlying graph reveals interesting phases of matter (\Fig{fig:phasediagram}). In the remainder of our study, we set $a_B=1$, $R=1.5$, and $r_{\tn{min}}=0.8$, unless stated otherwise.

To analyze the fate of the Hamiltonian at large $U$, we use the parton construction for the electronic operators, $c_{i,\alpha} = b_i f_{i,\alpha}$, in terms of a bosonic, electrically charged `rotor' field ($b_i=e^{i\theta_i}$) and a neutral fermion ($f_{i,\alpha}$) \cite{Florens_PRB_2004,Zhao_PRB_2007}. However, given the amorphous non-translationally invariant connectivity, we employ a new implementation of the two-site cluster mean field theory to obtain the phase-diagram. The method does not necessarily capture details of the energetics associated with the true many-body ground-state, but provides a self-consistent solution within the parton mean-field approximation. We will refer to the fermionic excitations, $f_{i,\alpha}$, as `spinons' even though they are not tied to any underlying fractionalization associated with an actual spin degree of freedom.

In terms of the new fields, the Hamiltonian is given by
\bea
H_0 &=& -\sum_{\substack{ i \neq  j \\ \alpha \beta }} \Big[   t_{\alpha\beta}(\br_{ij}) f_{i,\alpha}^\dagger b^\dagger_i b_j f_{j,\beta} + \tn{h.c.} \Big] + \sum_{i, \alpha, \beta} \epsilon_{\alpha \beta} f^\dagger_{i,\alpha} f^{\phantom\dagger}_{i,\beta},\nonumber \\
H_{\tn{int}} &=& \frac{U}{2}\sum_i (\hat{L}_{i}-1)(\hat{L}_{i}-2)
\label{eq:Ham}
\eea
where we have used the physical Hilbert space constraint  $L_i + \sum_\alpha n^f_{i,\alpha} = 2$ with $\hat{L}_i = \frac{\partial}{\partial \theta_i}$ and $\langle N_i\rangle = \sum_{\alpha}\langle n^f_{i,\alpha}\rangle$. We include a chemical potential to maintain half-filling of electrons which imposes  $ \langle \sum_\alpha n^f_{i,\alpha} \rangle=1$ and  $\langle L_{i}\rangle =1$ at every site $i$ for both spinons and rotors. We carry out a mean-field decoupling of $H_0\rightarrow H_b + H_f$, where nominally $H_f$ is described by a free Hamiltonian \cite{SI}. However, the fermion hoppings are renormalized by the rotor correlators, $\langle b_i^\dagger b_j\rangle$. As noted earlier, to solve for the interacting rotor-Hamiltonian, $H_b+H_{\tn{int}}$, we use a two-site cluster mean field theory for {\it every} pair of connected sites on the amorphous graph. We calculate the superfluid order parameter on every site $\langle b_i \rangle $, and the fermionic correlators between pairs of sites $\langle f^\dagger_{i,\alpha} f_{j,\beta} \rangle$ self-consistently until the results converge. 

\begin{figure}
\centering
\includegraphics[width=0.493\linewidth]{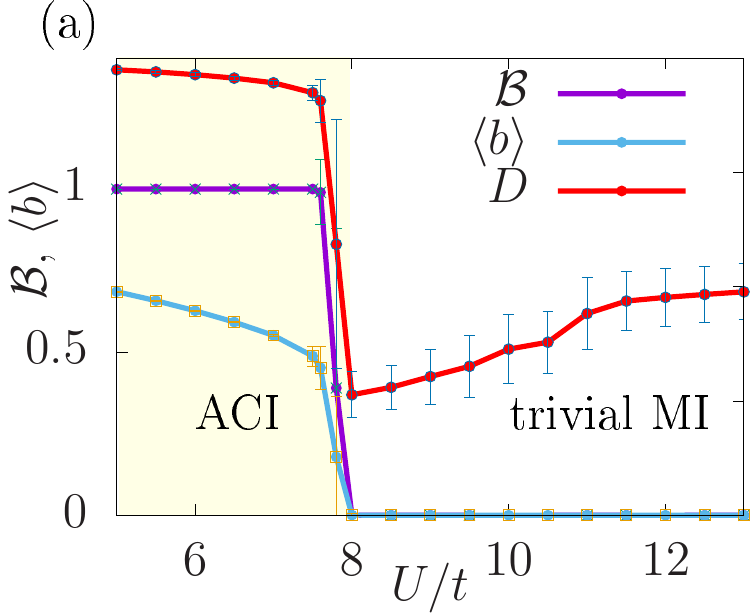}
\includegraphics[width=0.493\linewidth]{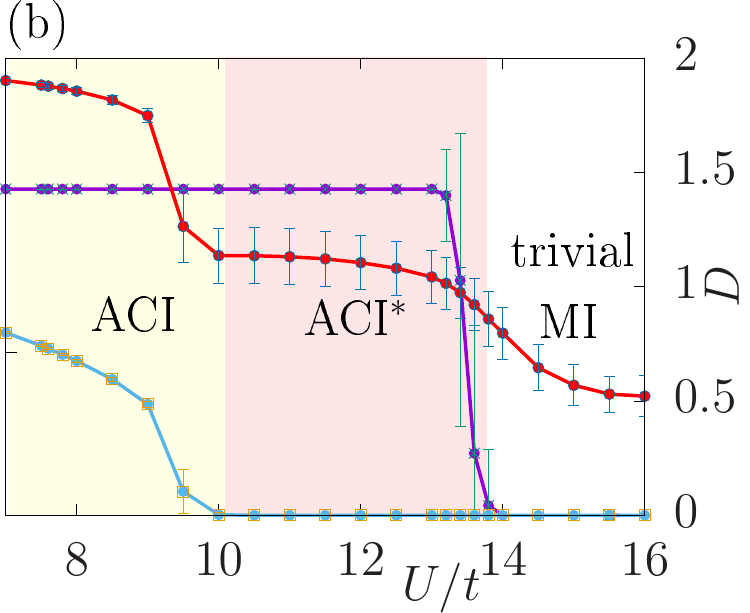}
\caption{Bott index ($\B$), superfluid order parameter $\langle b \rangle$, and the rotor fractal dimension $D$ in system of size $16\times 16$ for a fixed (a) $M/t=-1.84$, and (b) $M/t=-1.94$, respectively, obtained by averaging over 100 uncorrelated amorphous configurations.
}
\label{transition}
\end{figure}

{\it Fractionalized phases.-} Our solution suggests that for $U\lesssim 9t$, the rotors are condensed in a superfluid phase with $\langle b \rangle  = \sum_i \langle b_i \rangle/N~(\sim O(1))$, which is consistent with the typical values even in non-topological and crystalline systems \cite{Florens_PRB_2004,Zhao_PRB_2007}.  As a function of $M/t$, the spinon ``bands" realize both topological and trivial phases, respectively (\Fig{fig:phasediagram}). A non-zero $\langle b \rangle\neq0$ corresponds to a confined phase with a renormalized electronic quasiparticle residue; the two phases described above correspond to a trivial electronic insulator and an amorphous electronic Chern insulator which are adiabatically connected to the $U=0$ limit \cite{Agarwala_PRL_2017}. 

To diagnose the topological properties for the spinons in the absence of translational invariance and a notion of a Brillouin zone, we evaluate their {\it Bott index} ($\B$) which quantifies the non-commutativity of the position operators when projected to the lower ``band" of the spinon system \cite{Hastings_AOP_2011,Agarwala_PRL_2017}. The index is given by
\beq
\B =\frac{1}{2\pi} \text{Im} \{ \text{tr} [\log(WUW^\dagger U^\dagger) ]\}
\eeq
where $U=P \exp{(i \Theta)} P$, $W = P \exp{(i \Phi)}P$ and $P=\sum_{m}|m\rangle\langle m |$ is the projector made of all occupied spinonic wavefunctions $|m\rangle$. $\Theta, \Phi$ are diagonal matrices whose elements correspond to the real space positions of every orbital $(x_{i\alpha},y_{i\alpha})$, but compactified into a torus $(2\pi x_{i\alpha}/L,2\pi y_{i\alpha}/L)$, where $L$ is the linear dimension of the system. $\B$ is thus a generalization of the usual Chern number, but in a setting where translational symmetry is completely absent.

At larger values of $U/t$, we find two new correlation-induced phases (\Fig{transition}(a)-(b)). Depending on the value of $M/t$, we find distinct insulating regimes. For example, for $M/t\sim -1.84$ and as a function of increasing $U/t$, the superfluid order $\langle b \rangle $ goes to zero across $U = U_c$ . Across the same transition, and within our numerical resolution, the Bott index jumps from $1$ to $0$. Interestingly, both of these changes across $U\sim U_c$ occur over any typical configuration, such that configuration averaged quantities ($\sim O(100)$ configurations) also reflect a relatively sharp transition. The latter phase is clearly a topologically trivial Mott insulator realized at half-filling due to strong correlation effect. Even while $\langle b \rangle$ goes to zero, there are rare regions in the system where superfluid (SF) order parameter is vanishingly small, yet finite, which results in intrinsic inhomogeneities in this system, which we discuss in detail later in the paper. In contrast, for $M/t\sim -1.94$, there is an intermediate phase over $U_{c1}<U<U_{c2}$ where the rotors are Mott insulating with $\langle b \rangle = 0$, while the Bott index remains pinned at $1$. This intermediate phase is a novel example of an electronic Mott insulator where the spinons inherit non-trivial topology in an amorphous setting. As we discuss below, when defined in an open system with boundaries, the phase is defined by protected electrically {\it neutral} chiral edge modes that can carry a finite energy. We dub this phase as the fractionalized {\it amorphous} Chern insulator (ACI$^*$). We note that the fractionalized phases appear near $M/t\sim -2$, which corresponds to the critical point of the non-interacting crystalline model. This can be understood by noting that the inter-site fermionic hopping gets renormalized by the rotor correlators at large $U/t$, which then effectively reduces the spinonic bandwidth around $M/t=-2$. The non-trivial features of the spinonic band are likely to manifest near a window of $M/t\sim -2$. While the phase occurs over a small window of parameter space, our calculations with increasing system size suggest that they survive even in the thermodynamic limit \cite{SI}. 

{\it Glassy physics.-} A key element in our study is the inherently disordered nature of the amorphous network, which can lead to ``glassy" physics and strong local inhomogeneities. At small $U/t$, we find that the disorder induced ``rare-region" effects are relatively suppressed in the rotor superfluid and associated electronic trivial or Chern insulators; the distribution of site-resolved local superfluid order does not develop strong inhomogeneity. However, with increasing $U/t$, the amorphous network develops strong local inhomogeneities in the vicinity of the Mott transition. To characterize this phenomenology, we compute the fractal dimension for the rotors as \cite{geissler2021finite},
\bea
D=\log_L \left(\frac{\sum_{i=1}^{N} |\langle b_i \rangle|^2}{\tn{max}_{i}|\langle b_i \rangle|^2}\right),
\label{eq:dphi}
\eea
where $i$ is the position index, $L$ is the linear dimension and $N$ is the total number of sites. $D$ is useful for characterizing the effective dimensionality of the largest SF patch in the system. We note that when the rotors are condensed and the superfluid order is relatively homogeneous across the system, $D$ approaches a value $\sim 2$. On the other hand, across the Mott transition and regardless of whether the resulting insulator is (non-)topological, we find that $D$ develops a characteristic kink showing the rotor SF-to-MI transition happens in a non-uniform way over the network. In contrast, for a crystalline system, a MI state when diagnosed on finite size lattices have an infinitesimally small but {\it uniform} SF order parameter which  implies $D$ remains $\sim 2$ \cite{SI}. Here, the characteristic kink in $D$ reflects the realization of a bose-glass phase. That the transition is thermodynamically stable is reflected in the systematic system size scaling \cite{SI}, where the kink remains even as system size is increased. Our discussion so far has focussed on the effects of inhomogeneities in the boson sector. Next we shift our attention to the fermions.

\begin{figure}[!h]
\centering
\hspace*{0.4cm}\includegraphics[width=0.95\linewidth]{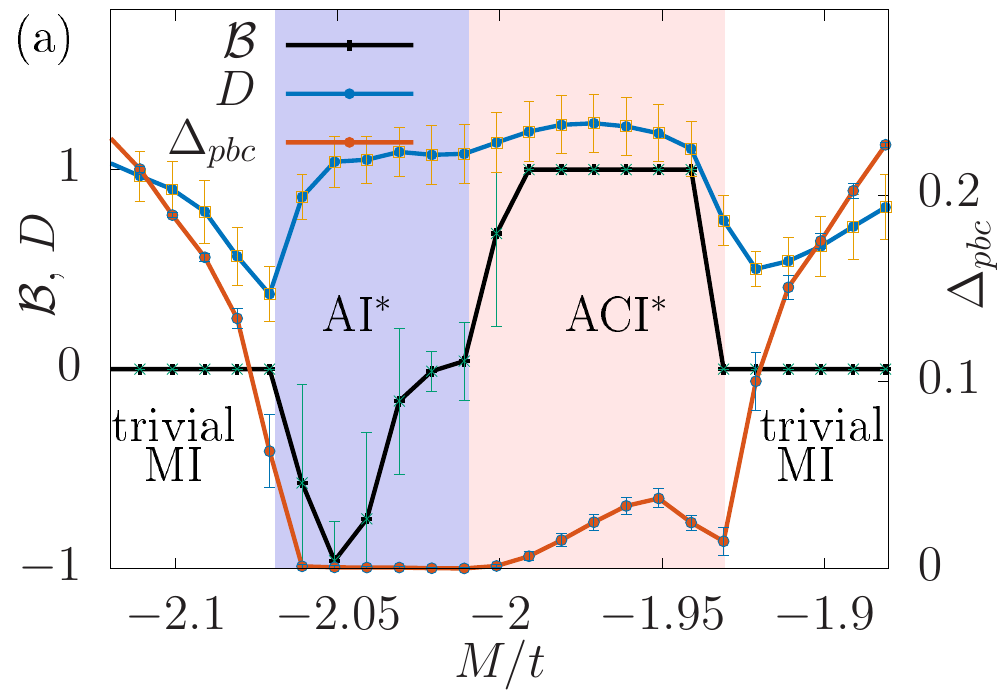}
\hspace*{-0.2cm}\includegraphics[width=0.88\linewidth]{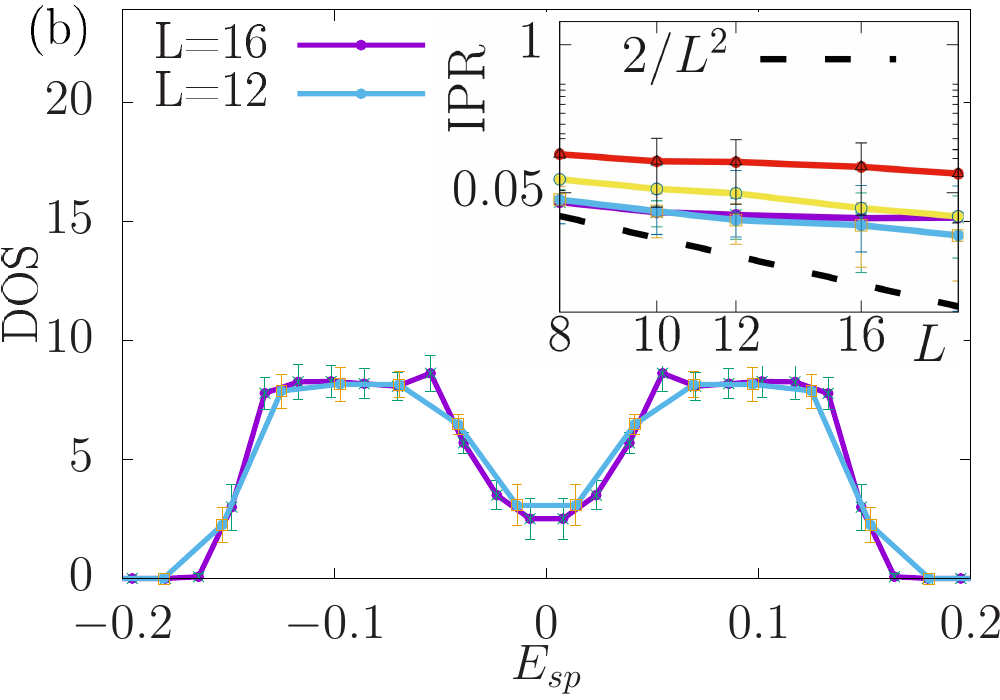}
\caption{(a) Configuration averaged quantities ($D, \B$, and $\Delta_{pbc}$) for a $16\times 16$ system at $U/t=12$ with periodic boundary conditions. (b) The spinon DOS in the AI$^*$ phase at $M/t=-2.05$, normalized by the total energy integrated DOS. Inset: The spinon IPR at the Fermi level for five different representative $M/t$ values in the AI$^*$ phase ($-2.05~\text{(violet)} \leq M/t \leq-2.02~\text{(red)}$) for increasing system size. The black dashed line marks the IPR of a homogeneous system.}
\label{fig:tbg}
\end{figure}

In the ACI$^*$ phase, the $f-$fermions have a finite gap, $\Delta_{\tn{pbc}}$, at half-filling (with $\langle  b \rangle =0$). Near $M/t \sim -2$, the spinon hopping competes with disorder energy scales mixing the two orbitals, leading to three distinct regimes (Fig.~\ref{fig:tbg}(a)): (i) for $M\lesssim -2.1t$  and $M \gtrsim -1.9t$,  $\Delta_{\tn{pbc}} \neq 0$ and $\B=0$, (ii) for $-2t \lesssim M \lesssim -1.9t$, $\Delta_{\tn{pbc}}\neq0$ but $\B=1$, and (iii) for $-2.1t \lesssim M \lesssim -2t$, $\Delta_{\tn{pbc}} \rightarrow 0$ while $\B\sim-1$ (see below). Over the entire range of $M/t$ values and large $U/t$ considered above, $\langle b \rangle \rightarrow 0 $ and $D<2$, which signifies that all of the phases are electrical insulators affected strongly by the inhomogeneity in the amorphous network. We have already discussed the origin of the interesting behavior in regions (i) and (ii) above. We note that these transitions are expected to remain sharp in the thermodynamic limit, as shown by the system size scaling \cite{SI}.

In region (iii), the ``gapless" phase ($\Delta_{\tn{pbc}} \rightarrow 0$) is not tied to any itinerant fermions, but arises instead due to localization on the amorphous network, as can be seen by calculating the inverse participation ratio (IPR),
\beq
\tn{IPR}_m =\frac{1}{2L^2}\sum_{i,\alpha}|\psi_{m,i\alpha}|^4,
\eeq
and the density of states (DOS) near the Fermi-energy. Here, $\psi_{m,i\alpha} = \langle i \alpha | m \rangle$ denotes the spinon wavefunction ($|m\rangle$) associated with orbital-$\alpha$ on the $i^{\tn{th}}$ site, and has been calculated from an explicit diagonalization of the mean-field Hamiltonian, $H_f$. We find a finite DOS at the Fermi level in the absence of a gap (see \Fig{fig:tbg}(b)); additionally, the IPR from the fermionic states is much larger than a typical delocalized state ($\sim 1/2L^2$) (inset of \Fig{fig:tbg}(b)). All of these observables are then consistent with a disorder induced localization of neutral spinons, and hence we dub it as a fractionalized Anderson insulator (AI$^*$). Interestingly, the seemingly non-trivial topological character associated with the large fluctuations of $\B$ near $-1$ is due to local topological ``puddles", associated with different disorder realizations.

\begin{figure}[h]
\centering
\hspace*{-0.05cm}\includegraphics[width=0.51\linewidth]{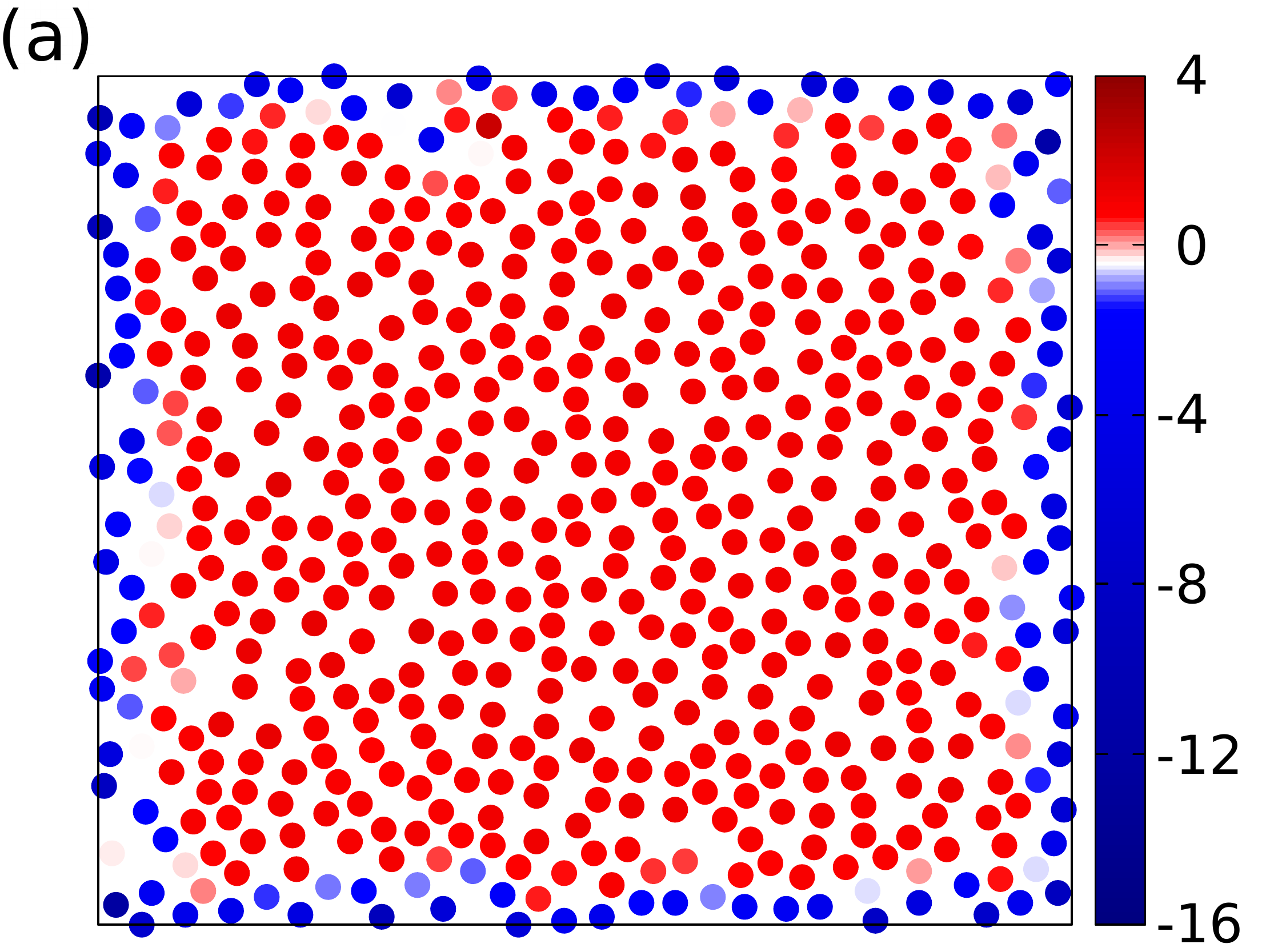}
\hspace*{-0.25cm}\includegraphics[width=0.51\linewidth]{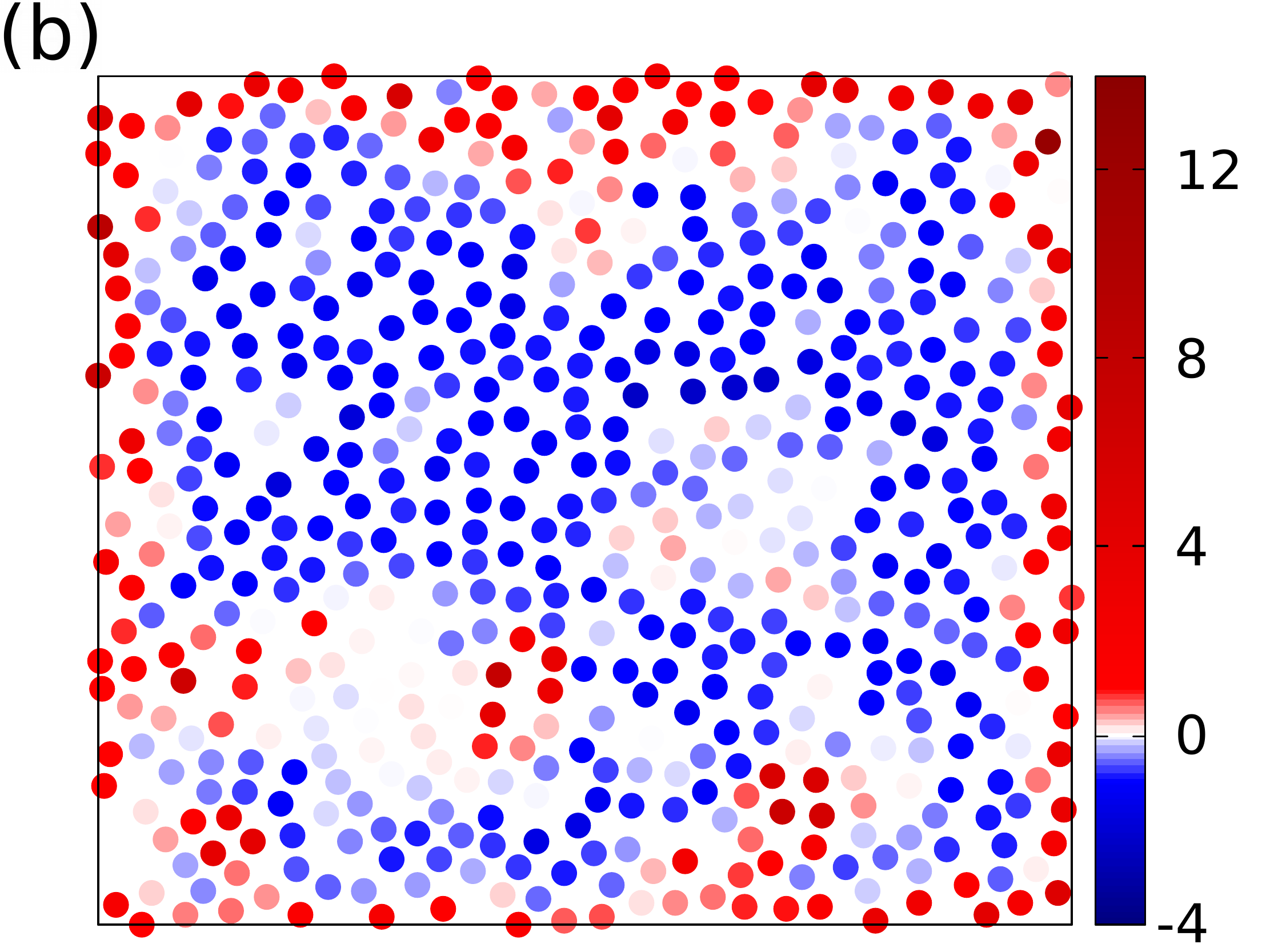}
\hspace*{-0.3cm}\includegraphics[width=0.47\linewidth,height=0.38\linewidth]{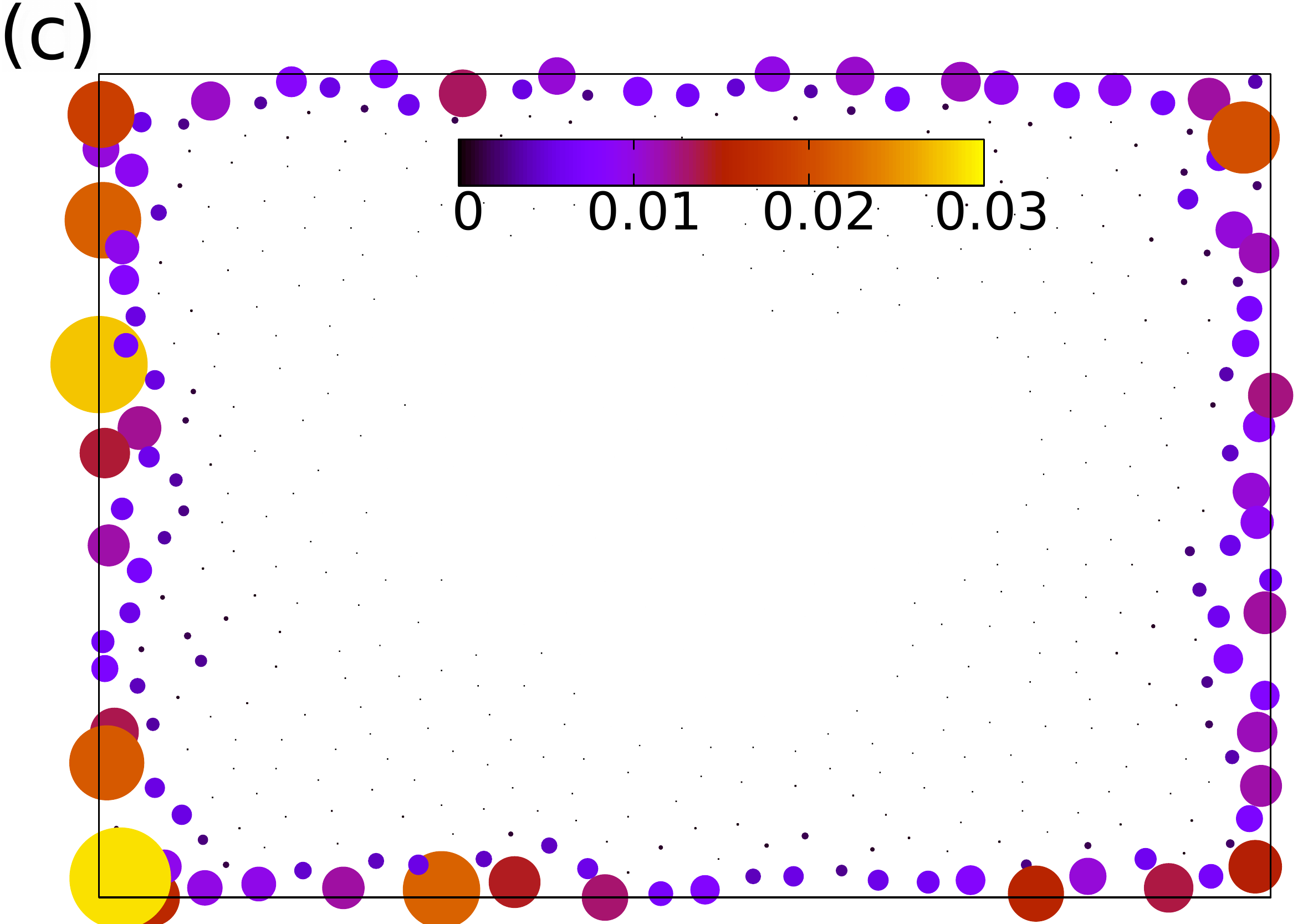}
\hspace*{0.1cm}\includegraphics[width=0.47\linewidth,height=0.38\linewidth]{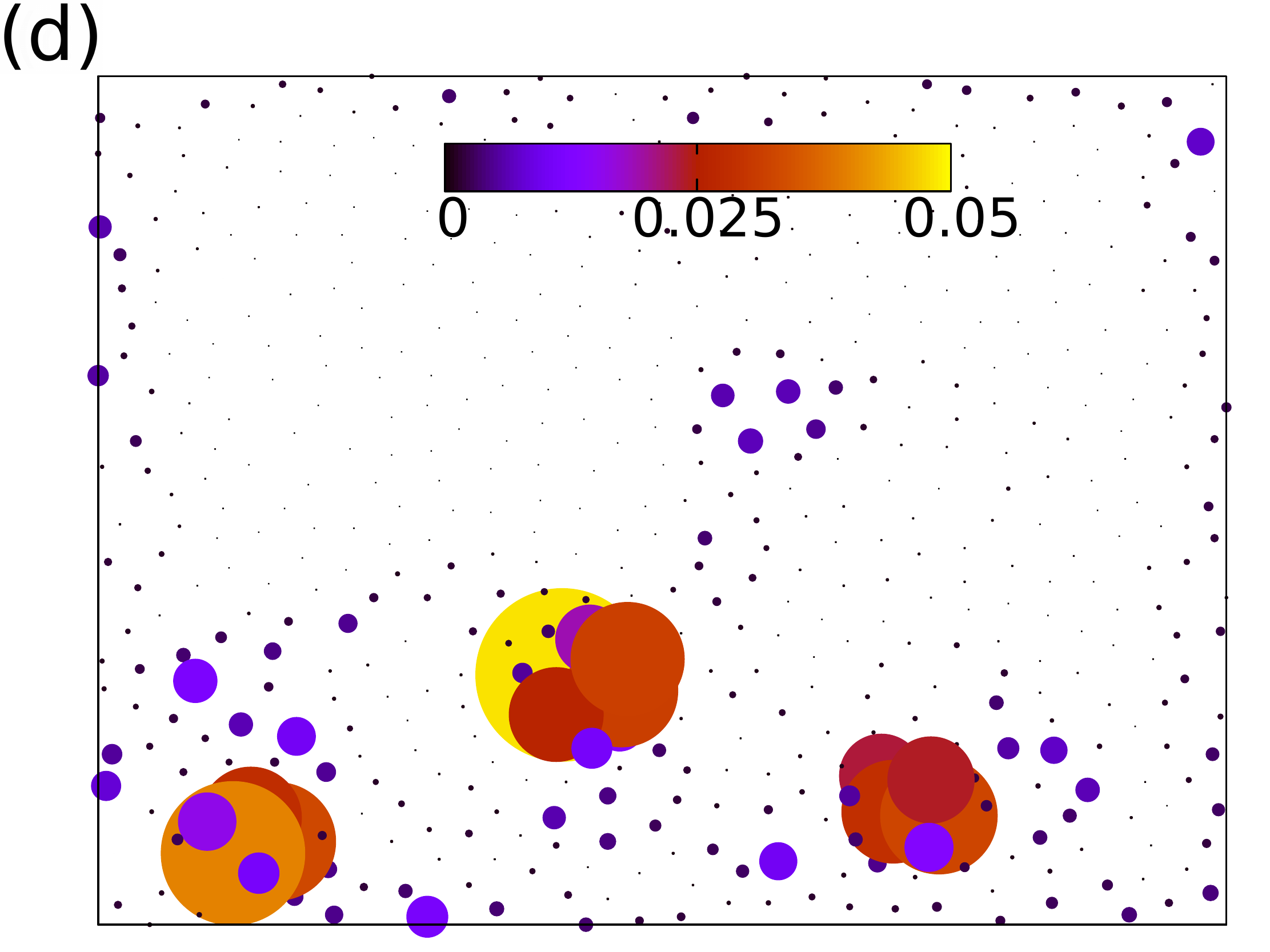}
\caption{The local Chern marker (top) and the midgap state
distribution (bottom) of the ACI$^*$ phase (left (a) and (c)) and the
AI$^*$ phase (right (b) and (d)) with open boundary conditions
(\(24\times 24\) system). For the midgap state distribution, the size and
the color of a blob quantify the probability of finding a spinon at
that site. In the ACI$^*$ phase, (a) and (c) show clear signs of a
gapless edge state.}
\label{local}
\end{figure}

Finally, to help clarify the topological character associated with the fractionalized phases, we also calculate a ``local Chern marker"
for a representative configuration in both the AI$^*$ and ACI$^*$ phases , respectively. The local Chern marker is defined as \cite{bianco2011},
\bea
C(\vec{r}) =-2\pi i \text{Tr}\{[PxP,PyP]\},
\label{eq:cmarker}
\eea
where $P$ is the projector, as defined earlier. The results for $C(\vec{r})$ for typical networks of size $24\times24$ are shown in Fig.~\ref{local} (a) and (b). In the ACI$^*$ phase, $C(\vec{r})$ is approximately uniformly distributed in the bulk (\Fig{local}(a)), while it exhibits stronger fluctuations in the AI$^*$ phase (\Fig{local}(b)) \cite{SI}. 
These observations are reflected in the location of the edge states in open systems. While the ACI$^*$ has spinon edge states (\Fig{local}(c)), the distinction between bulk and edge states is unclear in the AI$^*$ phase (\Fig{local}(d)). Thus even though $\B$ fluctuates strongly near $-1$ in the particular realization of the AI$^*$ phase in Fig.~\ref{local}(b), there is no clear sign of an edge state, and the midgap states are localized sporadically throughout the system. In particular, often wavefunctions are localized inside the bulk but between two domains of differently oriented local Chern marker patches reflecting development of edge like states but on effective grain boundaries. How such patches determine the physical characteristics of a thermodynamic phase and its properties is an interesting direction for future work. 

It is interesting to note that the disordered AI$^*$ phase only occurs in the $\mathcal{B}=-1$ regime. This has to do with the underlying non-interacting dispersion in the crystalline case, where the topological transitions are accompanied by Dirac cone closings at the $\Gamma$-point, $(0,\pi)$, $(\pi,0)$, and $(\pi,\pi)$, respectively. The transition to the $C=1$ phase arises at the $\Gamma$-point, which is more resilient in the amorphous setting. On the other hand, the transitions to the $C=-1$ phase arises at the large momentum points, which are more susceptible to the inhomogeneity associated with the amorphous network. 

{\it Outlook.-} 
In this work, we have analyzed the effects of strong correlations in the presence of non-trivial topology in an amorphous setting at the level of parton mean-field theory. None of the excitations we have obtained in the insulating bulk of these phases are itinerant. By including additional local degrees of freedom and by extending the analysis to higher dimensions \cite{Mukati_PRB_2020}, it will be interesting to analyze the possibility of hosting a diffusive spinon metal in a bulk electrical insulator coexisting with topological surface states. Going beyond the parton mean-field analysis, it is natural to address the effects of emergent gauge-field fluctuations, which are a defining feature of the fractionalized phases \cite{Ioffe-Larkin,Lee-Nagaosa_Wen,chowdhury2018mixed, Senthil_PRB_2008}. In our simplified construction of the ACI$^*$ phase, the total Bott index for the spinons is $\B=1$; in the analogous crystalline setup with Chern number $1$, the topological order and edge states can be gauged away \cite{wen2004quantum}. While we leave a detailed analysis of such effects for future work, a clear direction to explore further is to construct amorphous models with higher $\B$, where the edge states can not simply be gauged away. Similarly, in the AI$^*$ phase, the subtle interplay between disorder-induced localization and a tendency towards confinement in the presence of topological puddles is an interesting direction for future work. 
Finally, the platform considered here is ideally suited for studying the effects of non-trivial topology and fractionalization on their non-equilibrium glassy dynamics.

\acknowledgements We thank Chaoming Jian and Dan Mao for numerous insightful discussions and a critical reading of the manuscript. A.A. acknowledges many enlightening discussions and
related collaborations with Prateek Mukati, Bitan Roy, Subhro Bhattacharjee and Vijay B. Shenoy. SK and DC are supported by Grant No. 2020213 from the United States Israel Binational Science Foundation (BSF), Jerusalem, Israel. A.A. acknowledges partial financial support through Max Planck Partner group on strongly correlated systems at ICTS, and support from IIT Kanpur Initiation Grant (IITK/PHY/2022010).

\bibliographystyle{apsrev4-1_custom}
\bibliography{amorph}

\clearpage

\begin{widetext}

\setcounter{page}{1} 
\renewcommand{\figurename}{Supplemental Figure}
\renewcommand{\theequation}{\thesection.\arabic{equation}}


\begin{center}
    {\bf Supplementary material for ``Fractionalization and topology in amorphous electronic solids"}
\end{center}

\section{Additional details of the mean-field decomposition}

We present additional details of the mean-field decomposition in this section. Upon rewriting the Hamiltonian in terms of the spinonic and rotor fields, we can decouple the quartic term as

\bea
f_{i,\alpha}^\dagger f_{j,\beta}  b^\dagger_i b_j \rightarrow ^{\textbf{MFT}} \langle f_{i,\alpha}^\dagger f_{j,\beta} \rangle  b^\dagger_i b_j +  f_{i,\alpha}^\dagger f_{j,\beta} \langle   b^\dagger_i b_j \rangle 
- \langle f_{i,\alpha}^\dagger f_{j,\beta} \rangle  \langle   b^\dagger_i b_j \rangle .
\eea
Defining $\chi^{\alpha,\beta}_{ij} = \langle  f_{i,\alpha}^\dagger f_{j,\beta} \rangle$ and 
$\langle   b^\dagger_i b_j \rangle = B_{ij}$, we obtain mean-field Hamiltonians for each sector

\bea
\label{eq:mf_decomposition}
H^{b}_{\textbf{MFT}} &=&  - \sum_{ij}  \Big( \sum_{\alpha \beta} \chi^{\alpha,\beta}_{ij} t_{\alpha\beta}(\br_{ij}) \Big) b^\dagger_i b_j + h.c. + \sum_i U\frac{({L}_{i}-1)({L}_{i}-2)}{2} - \mu_{\theta} \sum_i {L_{i}} , \nonumber \\
H^{f}_{\textbf{MFT}} &=& -\sum_{ i \alpha } \sum_{ j \beta } B_{ij} t_{\alpha\beta}(\br_{ij}) f_{i,\alpha}^\dagger f_{j,\beta} - \mu_f \sum_{i,\alpha} (n_{i \alpha}^{f}) , \nonumber \\
\eea
where $\mu_f, \mu_\theta$  implement the filling constraints. Notice that spinon Hamiltonian is quadratic in fermionic operators, which can be simply diagonalized for a given value of $B_{ij}$.

\section{ Phase diagram in the weakly-interacting regime}

\renewcommand{\thefigure}{S1}
\begin{figure}[h]
\centering
\includegraphics[width=0.49\linewidth]{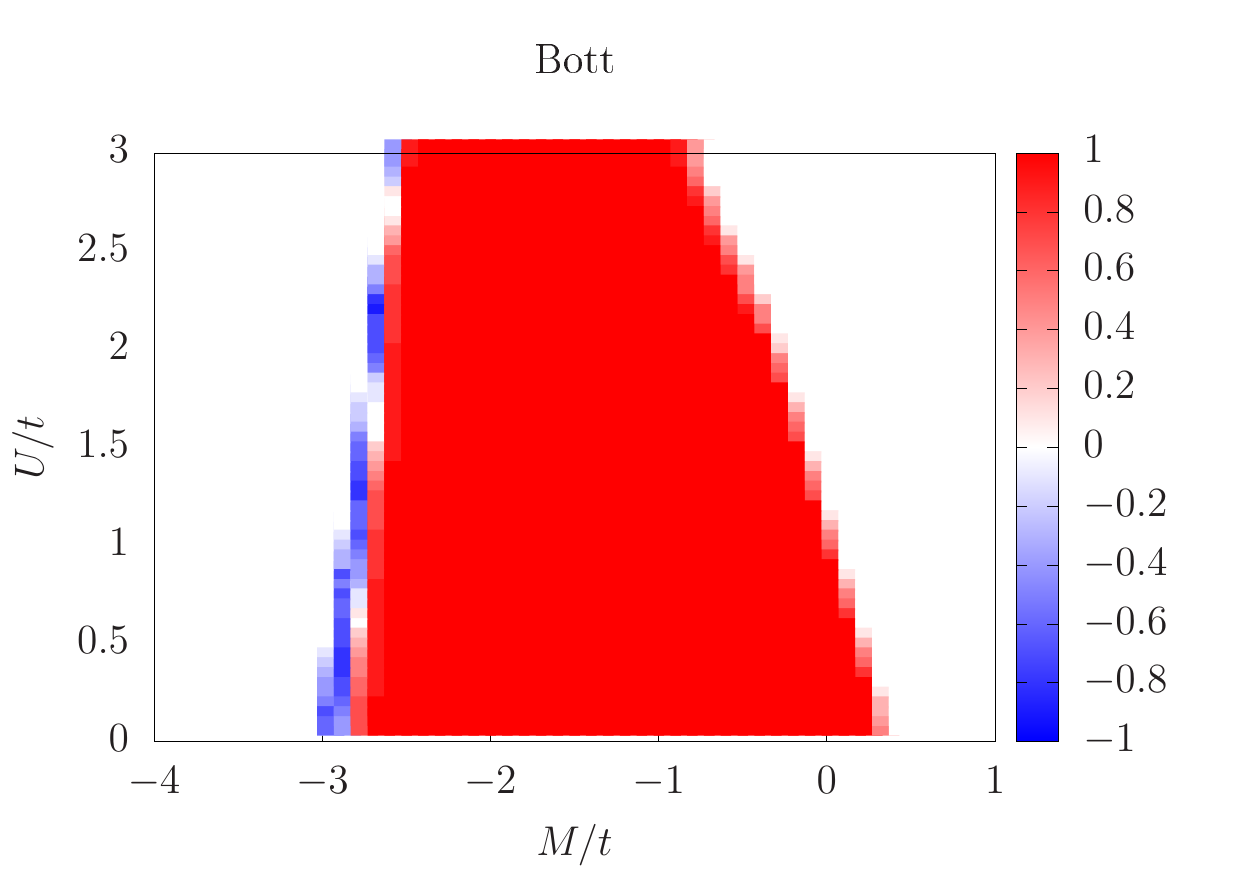}
\includegraphics[width=0.49\linewidth]{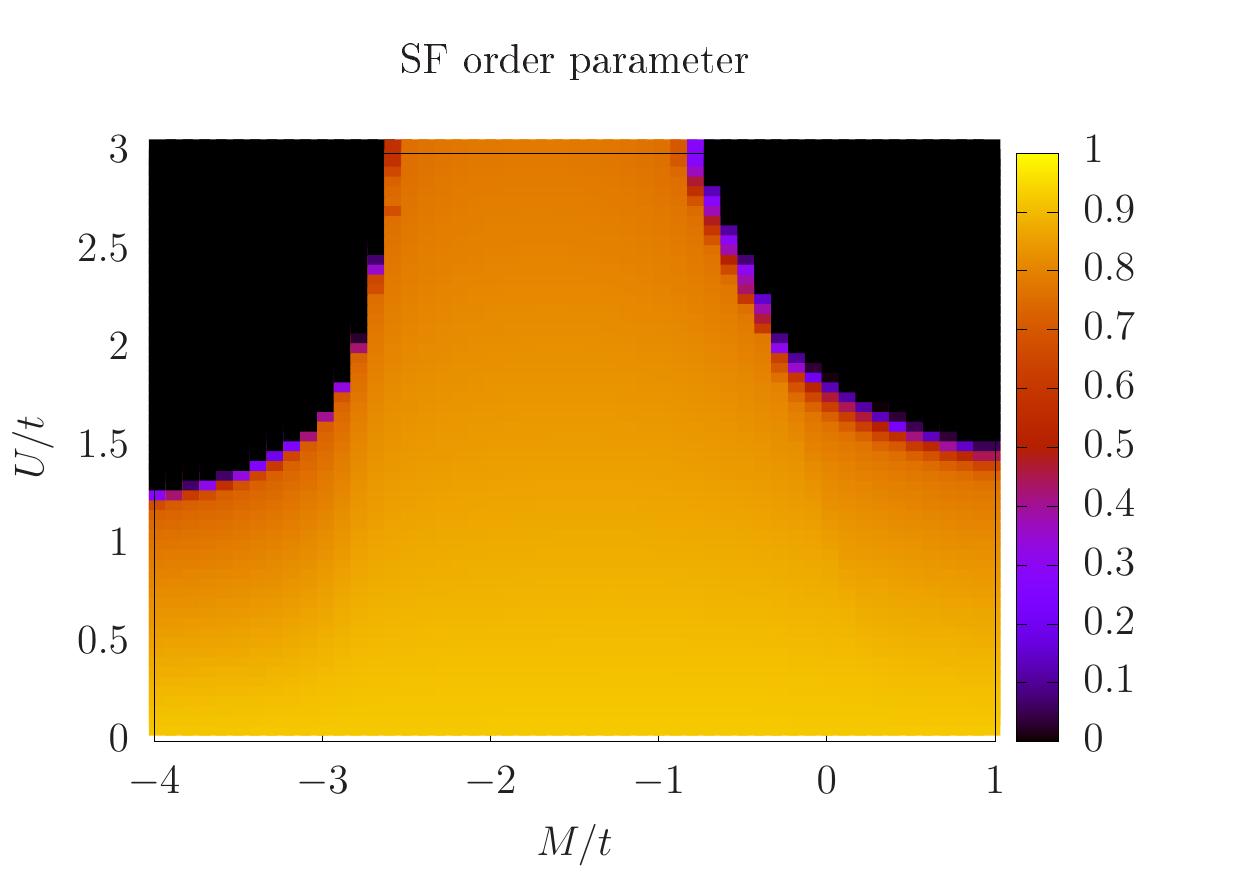}
\caption{Bott index and superfluid order parameter of $8\times 8$ systems in the weakly-interacting regime as a function of $M/t$ and $U/t$ (averaged over 50 configurations)}
\label{fig:appdx_lowu}
\end{figure}

In this section, we present the theoretical phase diagram of the amorphous systems in the weakly-interacting regime ($0\leq U/t\leq 3$). In the non-interacting limit, the system shows topological phase transitions as a function of $M/t$ between the ACI phase and the trivial band insulator phase, as shown in Fig.~\ref{fig:appdx_lowu}(a). As $U/t$ is increased, the range of $M/t$ hosting the topological phase becomes narrower. This can be readily understood by noting the decrease in $\langle b \rangle$ as a function of increasing $U/t$, as shown in Fig.~\ref{fig:appdx_lowu}(b). This change renormalizes the fermionic hopping coefficient (see Eq.~\ref{eq:mf_decomposition}), and thus reduces the range of the topologically non-trivial phase. We stress that in the weakly-interacting limit, the superfluid order parameter $\langle b \rangle$ has a finite value and the system hosts an \textit{electronic} phase: either the ACI phase or the band insulator phase.

\section{Bott index for non-interacting amorphous solids}
\label{sec:bott_rmin}

\renewcommand{\thefigure}{S2}
\begin{figure}[h]
\centering
\includegraphics[width=0.49\linewidth]{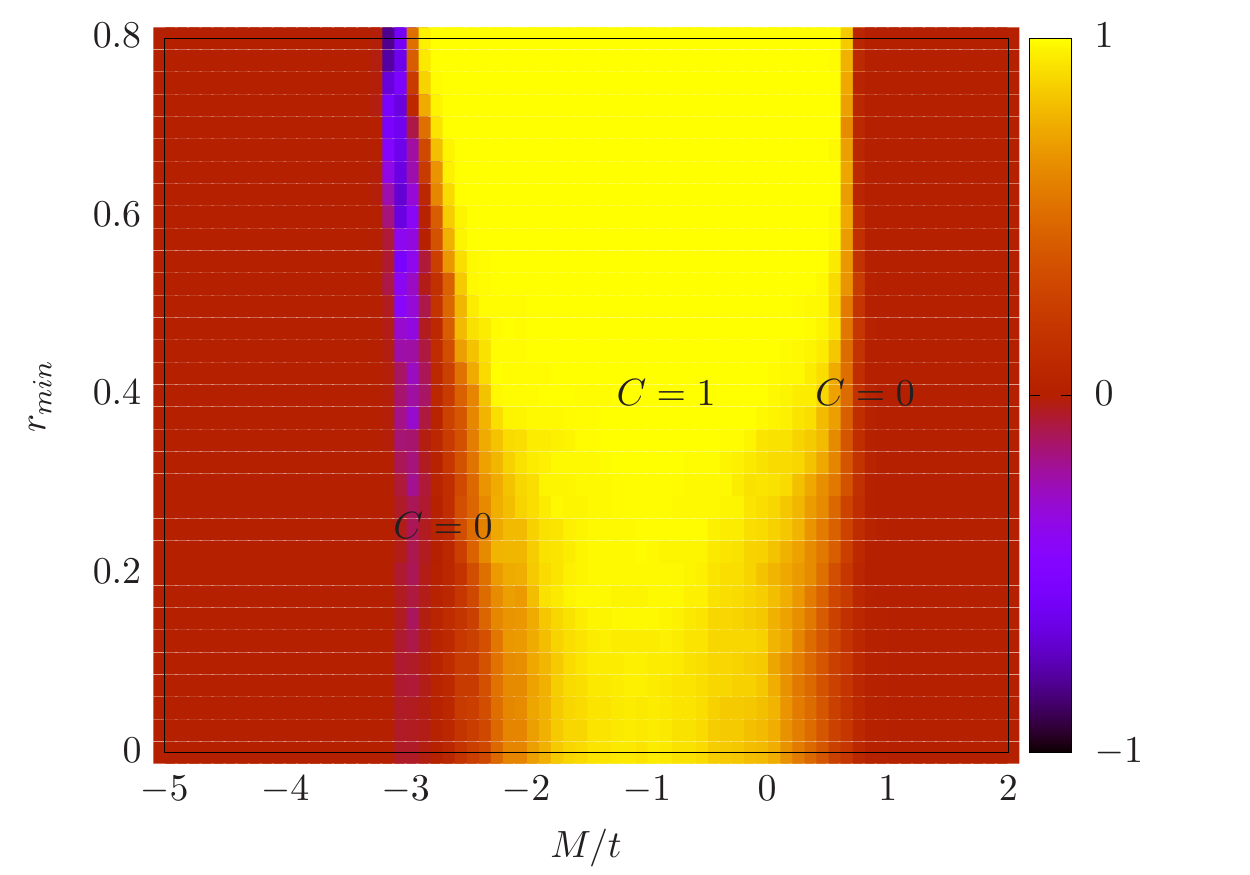}
\includegraphics[width=0.49\linewidth]{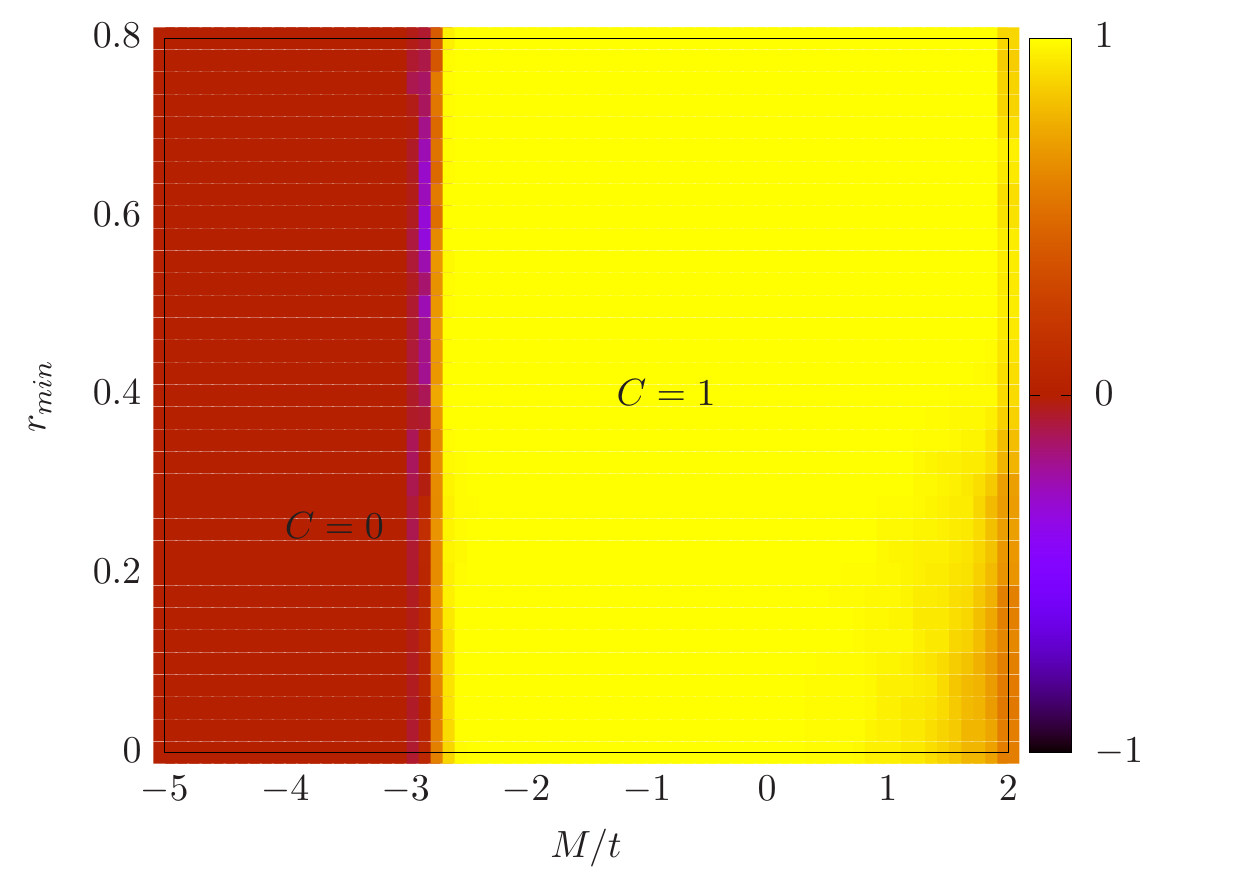}
\caption{Bott index of $8\times 8$ systems at $U=0$ as a function of $r_{\tn{min}}$ and $M/t$ (averaged over 100 configurations) for (a) $R=1.5$ and (b) $R=2$.}
\label{fig:appdx_Bott_rmin_M}
\end{figure}

We present the Bott index in the non-interacting limit as a function of $M/t$ and $r_\tn{min}$ with the cutoff distance $R=1.5$ and $R=2$ in Fig.~\ref{fig:appdx_Bott_rmin_M}. The perfect amorphous case corresponds to $r_{\tn min}=0$, in which the pair correlation function is uniform as a function of distance; we focused on the case of $r_\tn{min}=0.8$ in the main text. These results suggest that a free fermion amorphous system hosts a stable topological phase over a wide range of $r_\tn{min}$.

\section{Additional results as a function of system size}
\label{systemscaling}

\renewcommand{\thefigure}{S3}
\begin{figure}[h]
    \centering
    \includegraphics[width=0.49\linewidth]{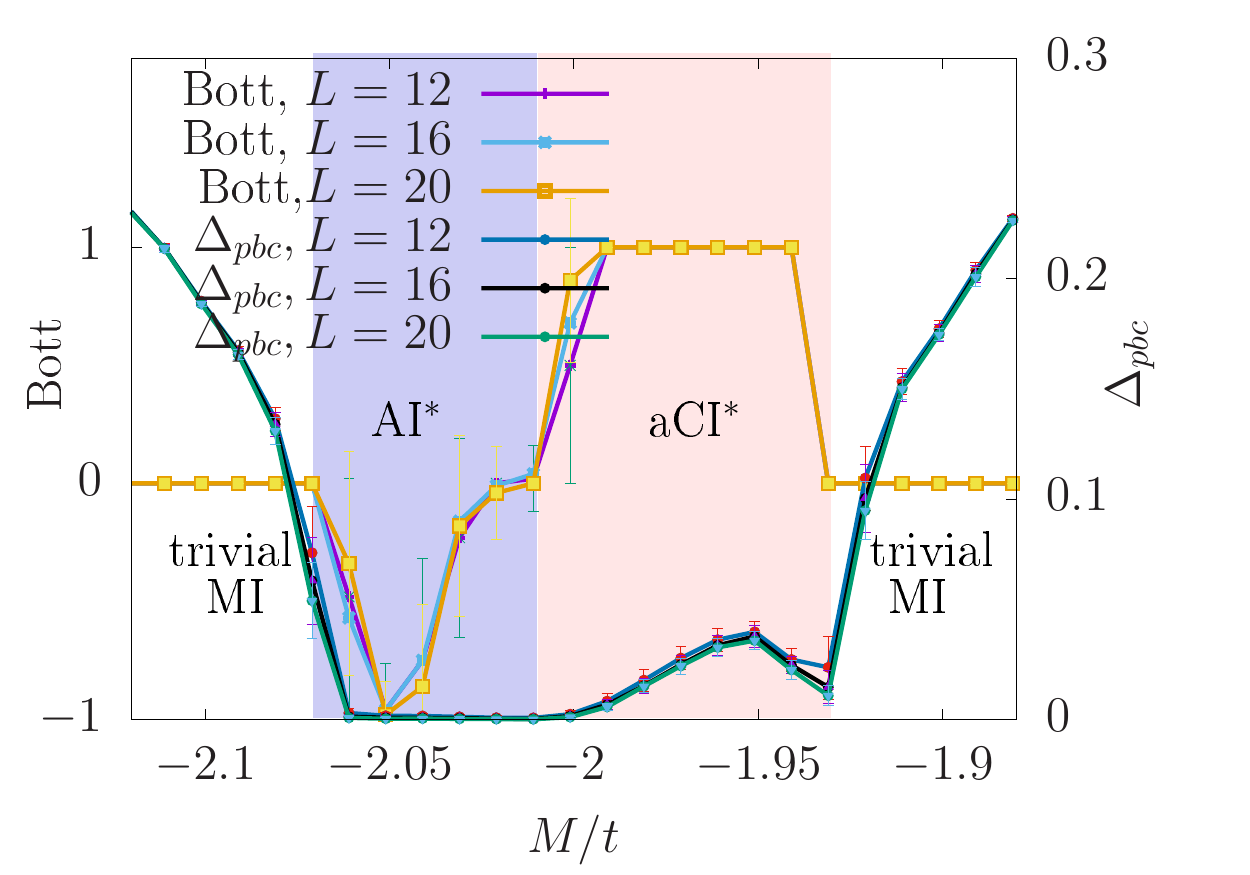}
    \includegraphics[width=0.49\linewidth]{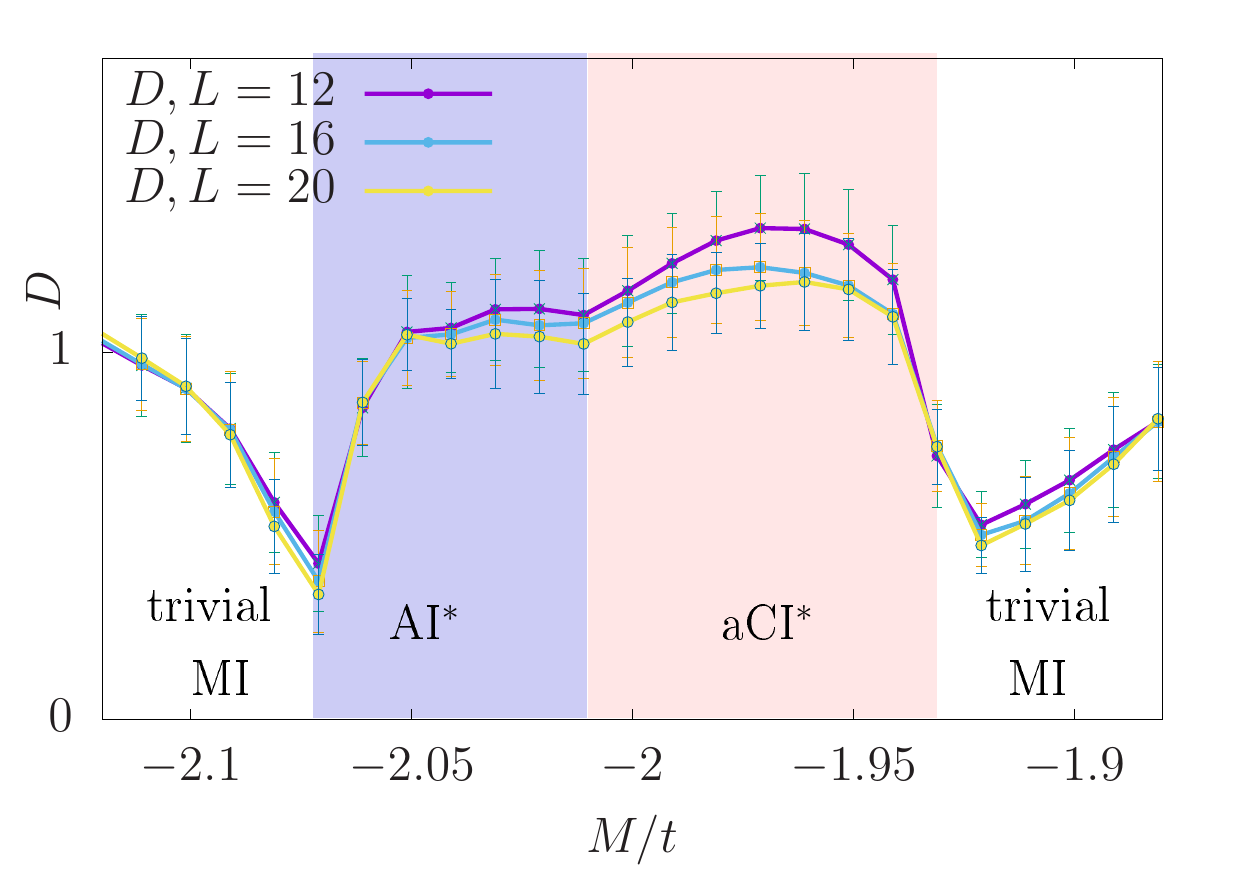}
    \caption{System size dependence of configuration averaged physical properties $(D, \B,$ and $\Delta_{pbc})$ for $U/t=12$ with periodic boundary conditions.}
    \label{fig:finitesize}
\end{figure}

We present additional results at various system sizes $L=12, 16, 20$ in Fig.~\ref{fig:finitesize}. We find that the spinon gap and the plateau with $\mathcal{B}=1$ remains finite with increasing system size, indicating the robustness of the ACI$^*$ phase even in the thermodynamic limit.

\section{Chern marker distributions}
\label{sec:cmarker}

\renewcommand{\thefigure}{S4}
\begin{figure}[h]
    \centering
    \includegraphics[width=0.49\linewidth]{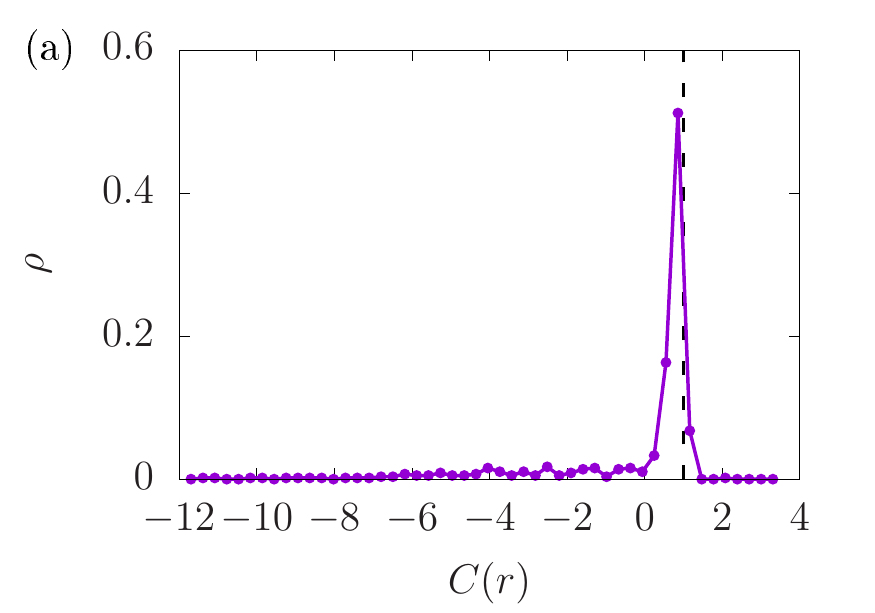}
    \includegraphics[width=0.49\linewidth]{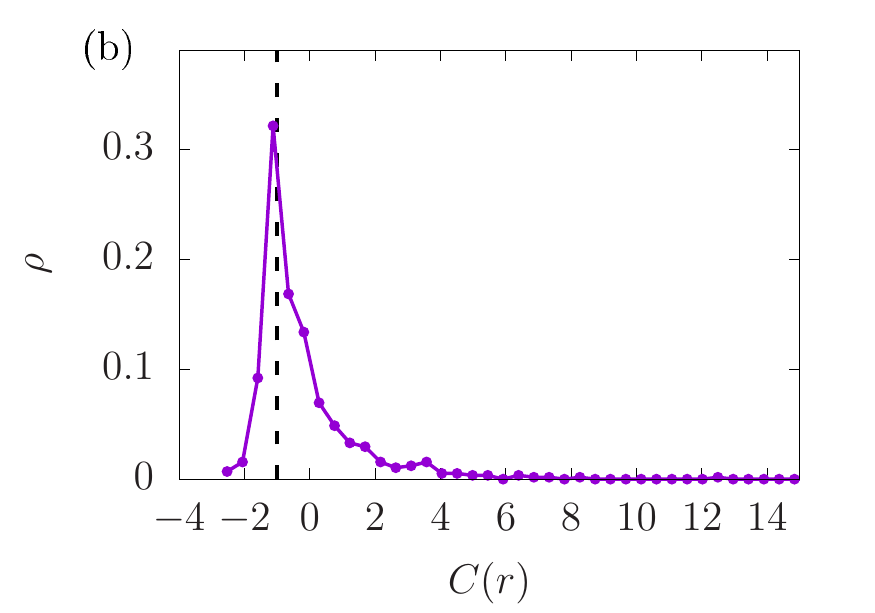}
    \caption{Chern marker distribution in (a) the ACI$^*$ phase, and (b) the AI$^*$ phase shown in Fig.~\ref{local} in the main text. The black dashed lines mark the Bott index values ($\pm1$) characterizing each phase under periodic boundary conditions.}
    \label{fig:cmarker_histo}
\end{figure}

We present the full distribution of Chern markers for the fractionalized phases addressed in Fig.~\ref{local}. For the ACI$^*$ phase (Fig.~\ref{fig:cmarker_histo}(a)), the Chern marker distribution shows a sharp peak near $C(r)=1$, pointing to an approximately homogeneous distribution of Chern marker values in the bulk. On the other hand, the AI$^*$ phase (Fig.~\ref{fig:cmarker_histo}(b)) shows a broad distribution between $-1\leq C(r) \lesssim 2$. This implies that the Chern marker exhibits a strong inhomogeneity even in the bulk. 

\section{Crystalline case}
\label{crystalcase}

\renewcommand{\thefigure}{S5}
\begin{figure}[h]
\centering
\includegraphics[width=0.45\linewidth, height=0.335\linewidth]{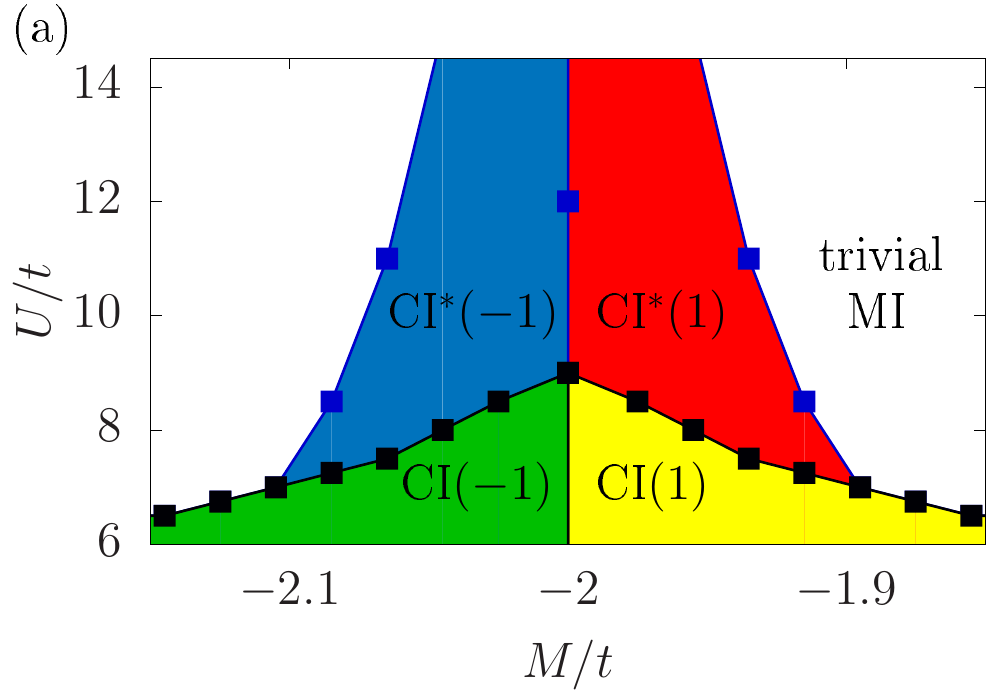}
\includegraphics[width=0.52\linewidth, height=0.34\linewidth]{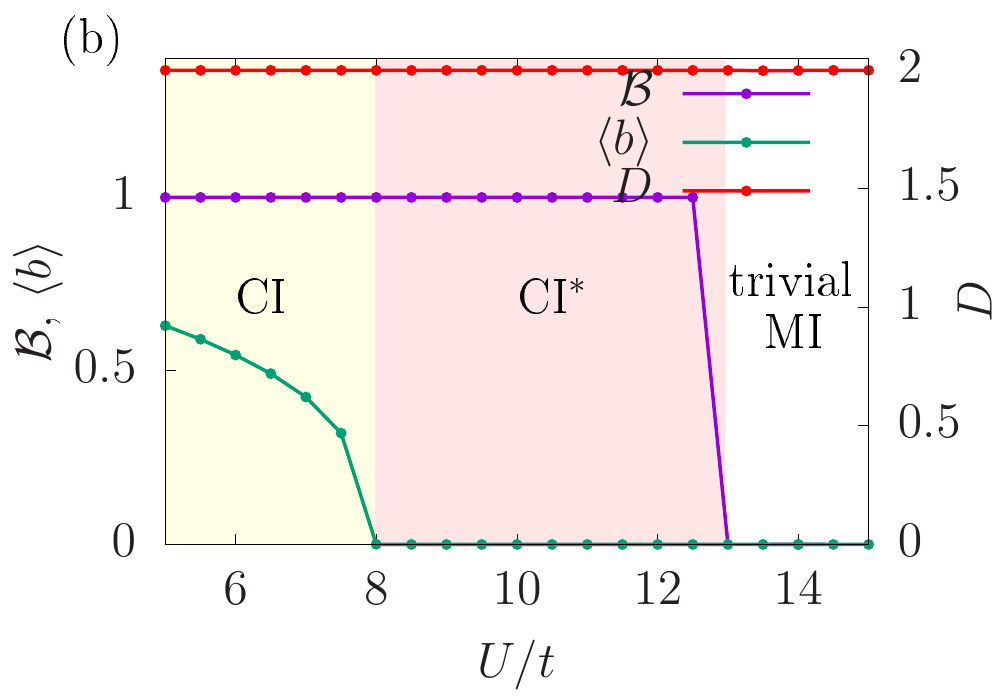}
\caption{Two site cluster results for the square lattice. (a) Phase diagram at half-filling as a function of $M/t$ and $U/t$. The four distinct phases denote a trivial electronic MI, an electronic CI, and CI$^*$ with $\mathcal{B}=\pm 1$. (b) $B, \langle b \rangle ,$ and $D$ as a function of increasing $U/t$ in system of size $16\times16$ at $M/t=-1.95$.} 
\label{fig:2cluster}
\end{figure}

We show the two-site cluster mean-field results on a square lattice with nearest neighbor hopping. There is a clear distinction between $\mathcal{B}=\pm1$ phases, and $D\sim 2$ throughout the whole phase diagram.

\newpage
\end{widetext}


\end{document}